\documentclass[preprint,12pt]{elsarticle}

\usepackage{graphicx}
\usepackage{graphics}
\usepackage{subfigure}
\usepackage{epstopdf}
\usepackage{harpoon}
\usepackage{amssymb}
\usepackage{amsmath}
\usepackage{float}
\usepackage{caption2}
\usepackage{geometry}
\usepackage{subfigure}
\usepackage{appendix}
\usepackage{mdwlist}

\journal{Journal}
\begin{document}

\begin{frontmatter}



\title{The invariant tori of knot type and the interlinked invariant tori in the Nos$\acute{\textup{e}}$-Hoover system}

\setcounter{footnote}{0}
\author{Lei Wang$^{1,2}$ and Xiao-Song Yang$^{1,}$\footnote{Author for correspondence\\\indent Email: xsyang@hust.edu.cn }}

\address{$^1$School of Mathematics and Statistics, Huazhong University of Science and Technology, Wuhan 430074, China\\
$^2$Department of Mathematics and Physics, Hefei University, Hefei 230601, China}

\begin{abstract}
  We revisit the famous Nos$\acute{\textup{e}}$-Hoover system in this paper and show the existence of some averagely conservative regions which are filled with an infinite sequence of nested tori. Depending on initial conditions, some invariant tori are of trefoil knot type, while the others are of trivial knot type. Moreover, we present a variety of interlinked invariant tori whose initial conditions are chosen from different averagely conservative regions and give all the interlinking numbers of those interlinked tori, showing that this quadratic system possesses so rich dynamic properties.
\end{abstract}


\end{frontmatter}


\section{Introduction}
In 1984, Shuichi Nos$\acute{\textup{e}}$ constructed a system called Nos$\acute{\textup{e}}$ equations to model the interaction of a particle with a heat-bath [1,2]. In 1986, Posch, Hoover and Vesely simplified the Nos$\acute{\textup{e}}$ equations by omitting an inessential variable and replaced the residual "momentum" by a "friction coefficient" and then got the following Nos$\acute{\textup{e}}$-Hoover equations [3]:
\begin{eqnarray}
\left\{\begin{array}{l}
\dot{x}=y\\
\dot{y}=-x-yz\\
\dot{z}=\alpha(y^2-1).\\
\end{array}\right.
\end{eqnarray}
where $\alpha$ is a positive real parameter. System (1) does not possess any fixed points and yet some interesting dynamics behaviors have been shown in [3]. For example, for some large value of parameter $\alpha$, the system exhibits both tori and chaotic trajectories, depending on initial conditions. \\
\indent Recently, Sprott et al. considered generalized Nos$\acute{\textup{e}}$-Hoover oscillators (the corresponding harmonic equations are $\dot{x}=y; \dot{y}=-x-zy; \dot{z}=y^2-1-\epsilon \textup{tanh}(x)$) [4]. They showed many interesting and surprising dynamic behaviors as follows. For some values of $\varepsilon$, they got that the conservative regions can coexist with dissipative regions in phase space by constructing appropriate cross-section. Especially, they showed that the systems owns interlinked invariant tori (see FIG 2 in [4]). Also, Sprott have found an other system without equilibrium that has a strange attractor and invariant tori [5]. However, it seems that the system studied in [5] does not possesses interlinked invariant tori.\\
\indent In this paper, we revisit the original Nos$\acute{\textup{e}}$-Hoover system (1) with parameter $\alpha=10$. By using proper cross-section, we show more types of invariant tori than those shown in [3], such as the invariant tori of trefoil knot type, the invariant tori with different types of symmetry, etc. Moreover, we show that the original system (1) possesses more complicated interlinked invariant tori than the generalized Nos$\acute{\textup{e}}$-Hoover oscillators shown in [4] in terms of interlinking numbers (see the following Section 3 for definition).
\section{The invariant tori of knot type and periodic orbits in six different averagely conservative regions }

Throughout the rest of this paper, we fix $\alpha=10$. When taking some initial conditions, the corresponding regular trajectories can trace out invariant tori in phase space. For example, for this initial condition $(x_{0},y_{0},z_{0})=(-0.72,0,0)$, an invariant torus is produced and its Lyapunov exponents are (0,0,0)(see Figure 1(a) and Figure 2).
In addition, chaotic trajectories can be found for some other initial conditions. For example, when taking the initial condition $(x_{0},y_{0},z_{0})=(-2.4,0,0)$, a chaotic trajectory with the Lyapunov exponents (0.0403 -0.0009 -0.0395) is produced (see Figure 1(b) and Figure 2). Numerical calculation shows that the chaotic trajectory is a so-called attractor. Since the system (1) has no equilibrium points, the attractor is "hidden" in the sense of Leonov and Kuznetsov [6-8], meaning that its basin does not intersect with small neighborhoods of any equilibrium, and it is cannot be found by standard computational methods.\\
\indent From Figure 1(c) and Figure 2, it can be seen that the invariant torus and the chaotic attractor are coexisting in phase space, which indicates the complexity of invariant sets of system (1).
 \begin{figure}[h]
\begin{center}
\subfigure[An invariant torus traced out by initial condition (-0.72,0,0).]{\includegraphics[width=4cm]{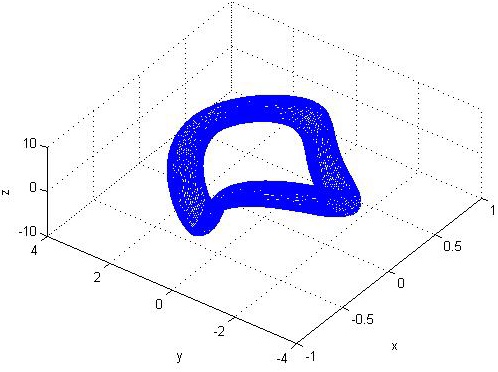}}
\hspace{0.06cm}
\subfigure[A chaotic attractor with initial condition (-2.4,0,0).]{\includegraphics[width=4cm]{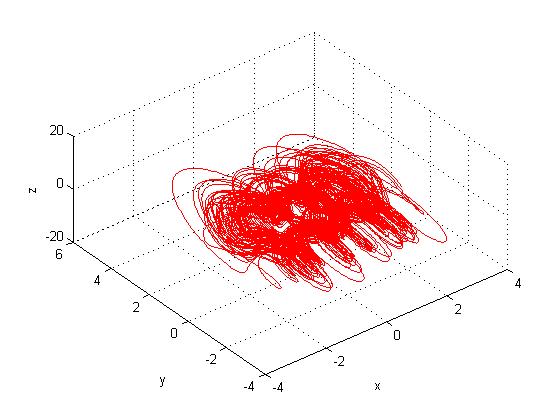}}
\hspace{0.06cm}
\subfigure[The interrelated locations of the torus in Figure 1(a) and the attractor in Figure 1(b).]{\includegraphics[width=4cm]{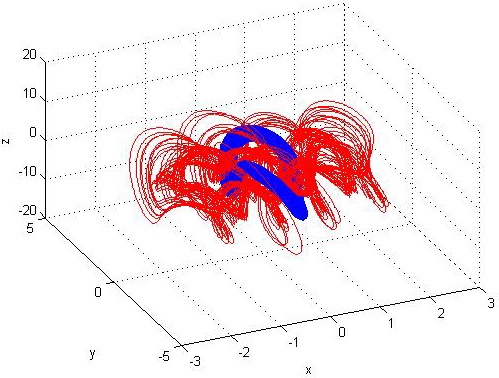}}
\end{center}
\caption{\small A torus and an attractor of system (1) with $\alpha=10$.}
\end{figure}
\begin{figure*}[h!t]
\centering
\includegraphics[height=6.5cm]{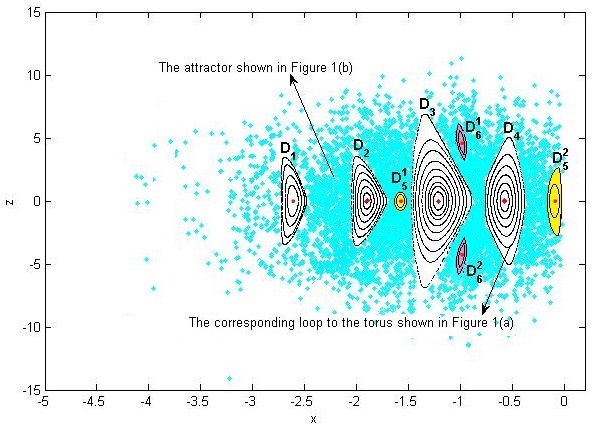}
\caption{\small Detailed cross-section image for $\alpha=10$ with 35 initial conditions taken over the interval $-4.5<x<0$ with $y=z=0$.
Here, white regions $D_{1},D_{2},D_{3},D_{4}$, yellow region $D_{5}=D_{5}^1\cup D_{5}^2$, and violet region $D_{6}=D_{6}^1\cup D_{6}^2$  are all filled with an infinite sequence of nested loops.}
\end{figure*}

 In order to demonstrate more clearly the complicated global structure of invariant sets of the system, we construct a cross-section $M=\{(x,y,z)\in \mathbb{R}^3|y=0\}$ as shown in Figure 2
 and study the Poincar$\acute{\textup{e}}$ map $P$ defined in the plane.
 Figure 2 shows that the six disjoint subsets  $D_{i}(i=1,2,...,6)$ contained in $M$ are filled with an infinite sequence of nested loops, where $D_{5}$ is the union of two yellow regions $D_{5}^1$ and $D_{5}^2$ , and $D_{6}$ is the union of two violet regions $D_{6}^1$ and $D_{6}^2$. Here, each loop is traced out by regular solution with different initial conditions. Moreover, we find that, for every initial condition in $D_{i}(i=1,2,3,4)$, the corresponding trajectory can generate a unique loop in $D_{i}$ under the iterated Poincar$\acute{\textup{e}}$ map . This seems to imply that those tori corresponding to the loops from $D_{i}(i=1,2,3,4)$ are of trivial knot type. The torus shown in Figure 1(a) above is just an example of trivial knot type and its initial condition is in $D_{3}$.  Nevertheless, for each initial condition in $D_{i}$ $(i=5,6)$,  the corresponding trajectory can generate two disjoint loops, one of which is in $D_{i}^1$ and the other is in $D_{i}^2$. This should indicate that the tori corresponding to the loops from $D_{5}$ and $D_{6}$ are of unusual knot type. For example, the following Figure 3 shows two invariant tori of trefoil knot type and their initial conditions are in $D_{5}$ and $D_{6}$, respectively.

\begin{figure}[h]
\begin{center}
\subfigure[An torus of trefoil knot type with the initial condition $(- 1.613,0,0)$ in $D_{5}$.]{\includegraphics[width=6cm]{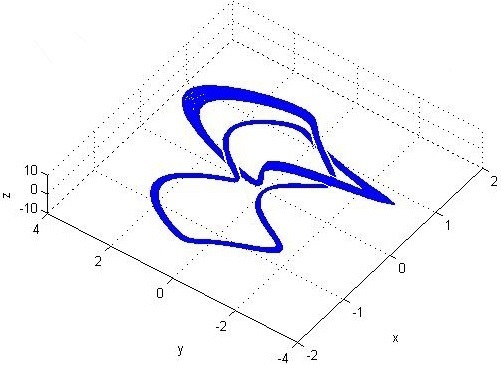}}
\hspace{0.06cm}
\subfigure[An torus of trefoil knot type with the initial condition (-1,0,-6) in $D_{6}$.]{\includegraphics[width=6cm]{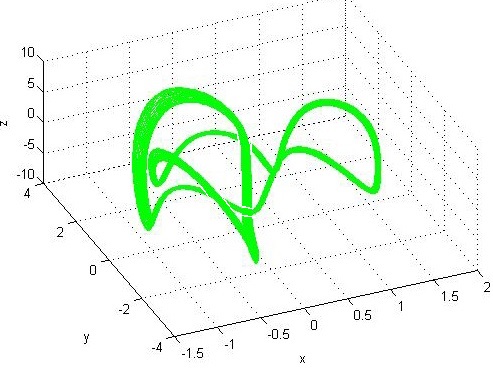}}
\end{center}
\caption{\small Two tori of trefoil knot type.}
\end{figure}
We note that not all regular trajectories can fill full of the surface of a torus. This is due to a well known result in theory of dynamical systems [9], i.e., for the orientation-preserving rotating map of circle, if the rotation number is rational, each orbit is periodic and can not fill full of the circle; Otherwise, each orbit can almost fill full of the circle if the rotation number is irrational. Thus, for the system (1), considering the Poincar$\acute{\textup{e}}$ map $P$ restricted to one of the loops in $M$, if the rotation number of this map is a rational number, one regular trajectory produces a finite string dots in the loop, which implies that we must choose several appropriate initial conditions to produce the closed loop in the cross-section. In contrast,  one regular trajectory produces a dense set of dots in the loop if the rotation number is irrational. \\
\begin{figure}[h]
\begin{center}
\subfigure[A period orbit with the initial condition in $D_{4}$. ]{\includegraphics[width=4.5cm]{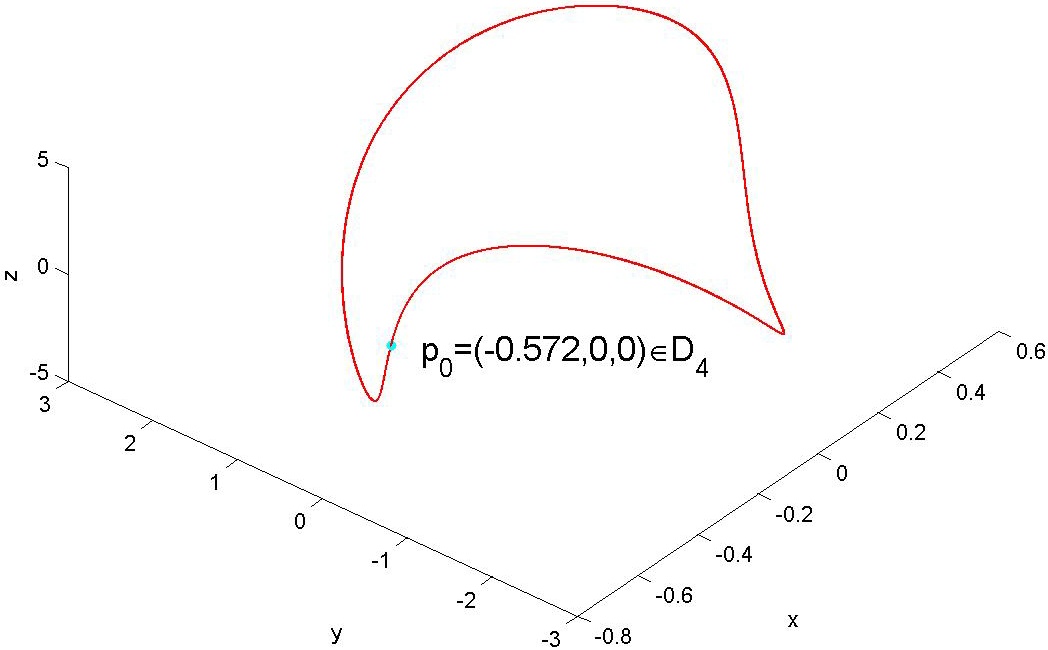}}
\hspace{0.06cm}
\subfigure[A period orbit of trefoil knot type with the initial condition in $D_{5}$.]{\includegraphics[width=4.5cm]{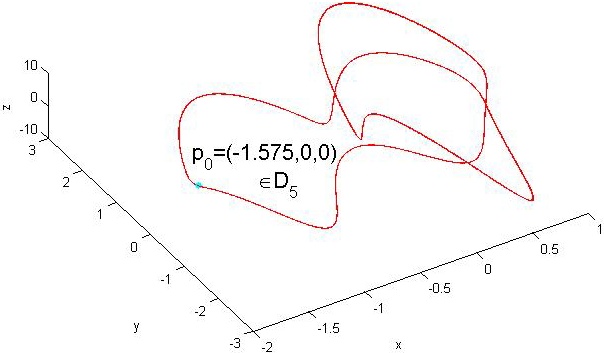}}
\end{center}
\caption{\small Two period orbits as examples.}
\end{figure}\\
 \indent As we all know, a fixed point of Poincar$\acute{\textup{e}}$ map corresponds to a periodic orbit of system (1). Since the first return Poincar$\acute{\textup{e}}$ map $P$ is a continuous mapping from $D_{i}$ to $D_{i}$($i=1,2,3,4$), there exists at least a fixed point in $D_{i}$ ($i=1,2,3,4$) for $P$ by Brouwer fixed point theorem. Nevertheless, for $i=5,6$, $P$ maps $D_{i}^1$ to $D_{i}^2$ and $D_{i}^2$ to $D_{i}^1$. Thus, the second return Poincar$\acute{\textup{e}}$ map $P^2$ map $D_{i}^1$ to $D_{i}^1$($i=5,6$),  which shows that there exists at least a fixed point in $D_{i}^1$($i=5,6$) for $P^2$ by Brouwer fixed point theorem. From the discussing above, we conclude that the system (1) has at least six periodic orbits. In Figure 2, the six small red dots in regions $D_{1},D_{2},D_{3},D_{4},D_{5}^1$ and $D_{6}^1$ indicate six different periodic orbits of system (1) respectively and two of them are shown in Figure 4 above as examples.\\
\indent Numerical calculation shows that, if randomly taking $x_{0}\in D_{i}(i=1,2,...,6)$ , we obtain all $\textup{det}(DP(x_{0}))\approx 1$ for $x_{0}\in D_{i}$ ($i=1,2,3,4$) and $\textup{det}(DP^2(x_{0}))\approx 1$ for  $x_{0}\in D_{i}$ ($i=5,6$)(See Appendix for the Matlab code). This certifies that, for any $i=1,2,3,4,5,6$, the phase space domain  $V_{i}=\bigcup\limits_{t=-\infty}^{\infty}\varphi_{t}(D_{i})$ which is filled with an infinite sequence of nested tori should obey time-averaged version of the Liouville's theorem [4]. Here, $\varphi_{t}(\cdot)$ is a flow generated by system (1). Thus we treat $V_{i}$ ($i=1,2,...,6$) as averagely conservative regions.
\begin{figure}[h]
\begin{center}
\subfigure[An invariant torus contained in $V_{4}$ which itself is symmetrical about $z$-axis.]{\includegraphics[width=5cm]{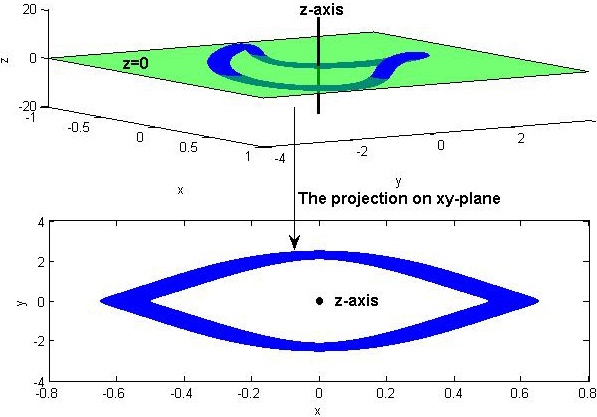}}
\hspace{0.06cm}
\subfigure[Two invariant tori with interlinking number 8 (see Section 3), which are symmetrical to each other about $z$-axis. Here the red torus is in $V_{6}$ and the blue torus is in $V_{5}$.]{\includegraphics[width=5cm]{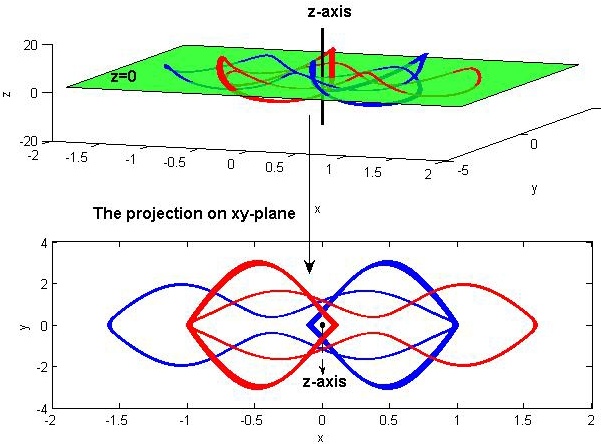}}
\end{center}
\caption{\small Two types of symmetry about $z$-axis.}
\end{figure}\\
\indent Obviously, system (1) is invariant under the transformation of coordinates $(x,y,z)\rightarrow(-x,-y,z)$, which indicates that, for any invariant torus of system (1), the torus generated by a 180$^\circ$ rotation of this invariant tori about $z$-axis is still an invariant torus of the system. Numerical simulation shows that, for system (1), there exists two types of symmetry of tori about $z$-axis. The first type is these tori contained in $V_{i}$$(i=1,2,3,4)$, any of which itself is symmetrical about $z$-axis, such as the torus shown in Figure 5(a) above. The second type is these tori contained in $V(5)$, any of which will be transformed into another invariant torus contained in $V(6)$ under the symmetry transformation about $z$-axis and vice versa, such as the two tori shown in Figure 5(b).\\
 \indent In addition, since $\nabla\cdot f=-z$, all of the tori in $V_{i}$ must go though plane \{$(x,y,z)|z=0$\} (see Figure 5).

%Moreover, if randomly taking sufficient initial conditions in $M-\bigcup_{i=1}^6D_{i}$,
 %numerical calculation shows that all trajectories will be attracted into the chaotic attractor $C$. Hence, the region $M-\bigcup_{i=1}^6D_{i}$ should be dissipative.
 %Thus, the averagely conservative region and the dissipative region are coexisting by Figure 2.
 %Furthermore, in phase-space, the locations of the solutions producing tori thread through the chaotic attractor.

\section{All interlinking numbers of the different interlinks of tori}
As shown in Subsection 2.1, for any $i=1,2,..,6$, the tori which stem from $D_{i}$ form an infinite nested sequence.
 However, the interrelated locations of tori derived from different averagely conservative regions are still not clear so far.
Now, by computer simulation, we show the interesting and magnificent interlocked phase-space structures in Figure 6.
Here, the six tori shown in Figure 6 are produced by utilizing the six initial conditions (-2.48,0,0), (-1.97,0,0), (-1.40,0,0), (-0.72,0,0), (-1.61,0,0) and (-1,0,-6)
which belong to $D_{1}$, $D_{2}$, $D_{3}$, $D_{4}$, $D_{5}$ and $D_{6}$ respectively, and all their Lyapunov exponents are (0,0,0).
\begin{figure}[h]
\begin{center}
\subfigure[$T_{12}$]{\includegraphics[width=3.0cm]{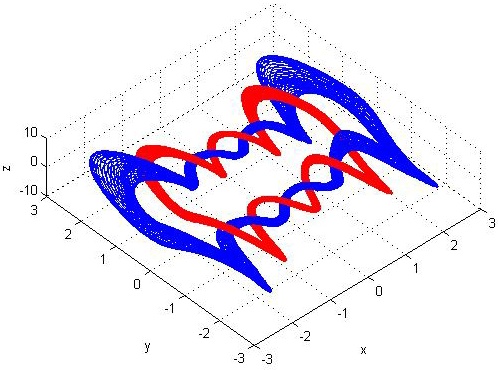}}
\subfigure[$T_{13}$]{\includegraphics[width=3.0cm]{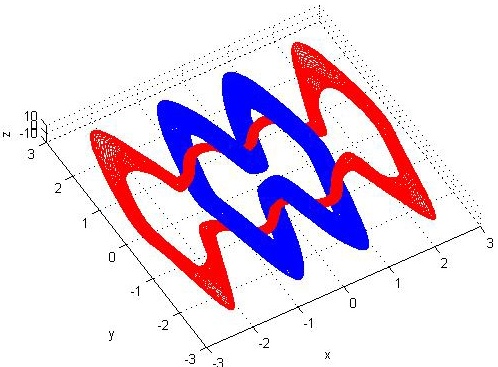}}
\subfigure[$T_{14}$]{\includegraphics[width=3.0cm]{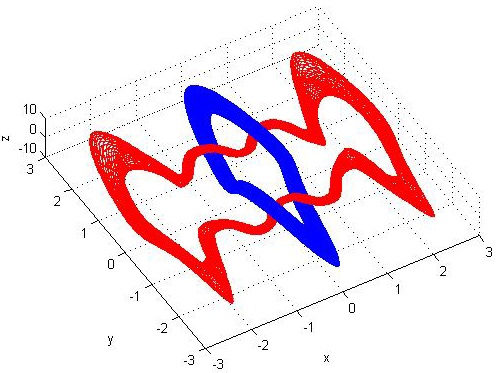}}
\subfigure[$T_{15}$]{\includegraphics[width=3.0cm]{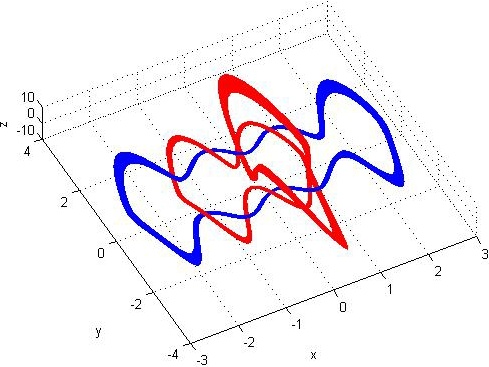}}
\subfigure[$T_{16}$]{\includegraphics[width=3.0cm]{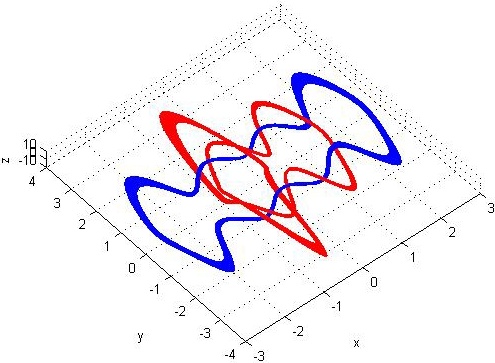}}
\subfigure[$T_{23}$]{\includegraphics[width=3.0cm]{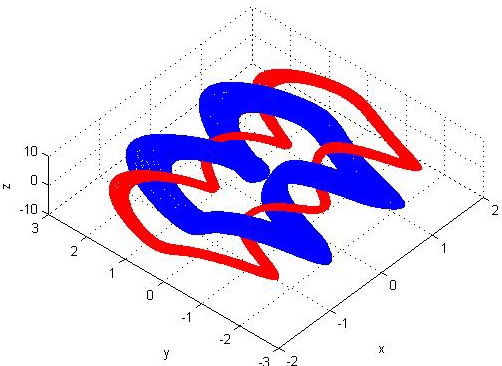}}
\subfigure[$T_{24}$]{\includegraphics[width=3.0cm]{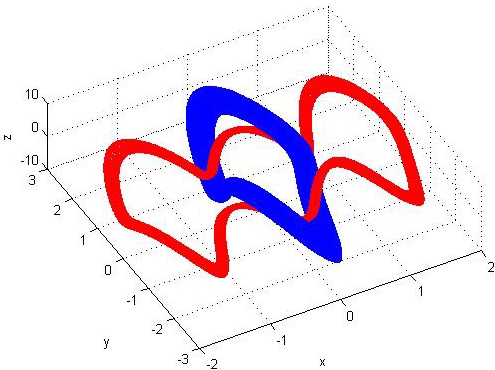}}
\subfigure[$T_{25}$]{\includegraphics[width=3.0cm]{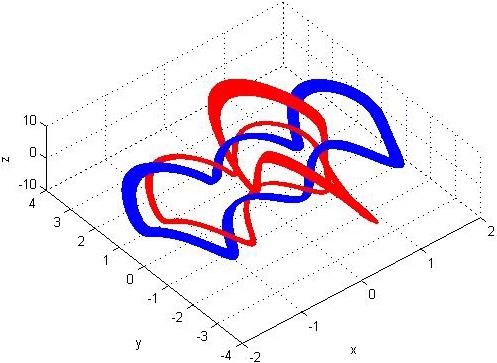}}
\subfigure[$T_{26}$]{\includegraphics[width=3.0cm]{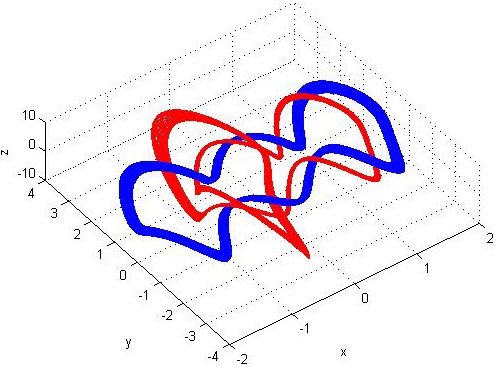}}
\subfigure[$T_{34}$]{\includegraphics[width=3.0cm]{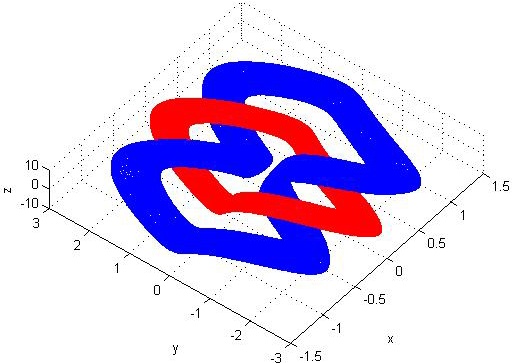}}
\subfigure[$T_{35}$]{\includegraphics[width=3.0cm]{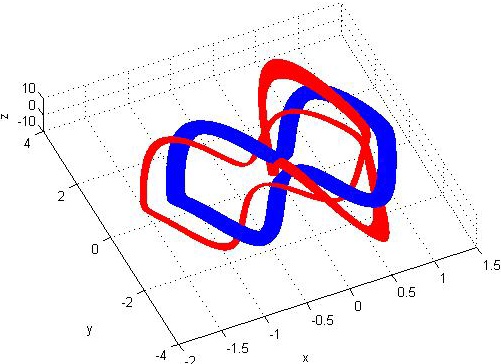}}
\subfigure[$T_{36}$]{\includegraphics[width=3.0cm]{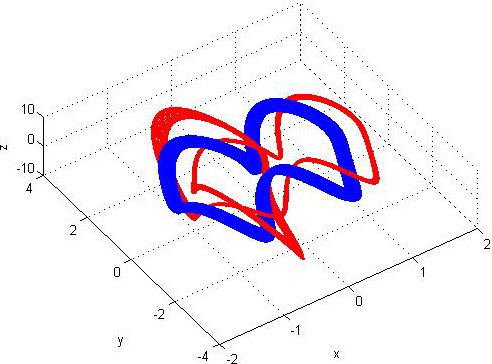}}
\subfigure[$T_{45}$]{\includegraphics[width=3.0cm]{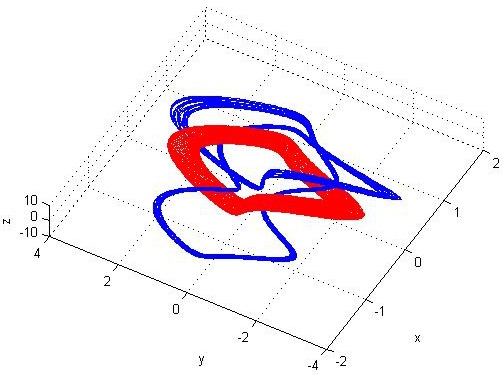}}
\subfigure[$T_{46}$]{\includegraphics[width=3.0cm]{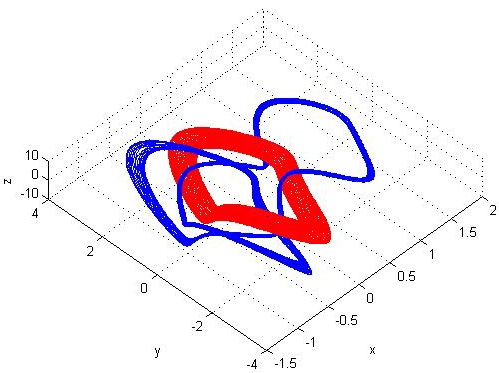}}
\subfigure[$T_{56}$]{\includegraphics[width=3.0cm]{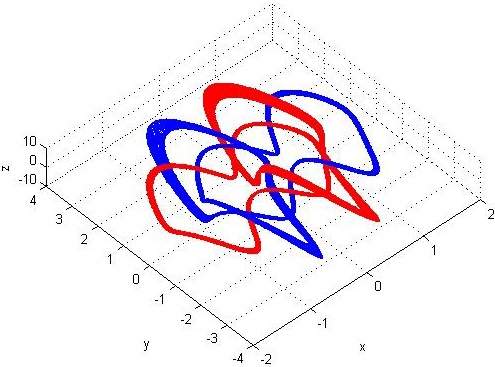}}
\end{center}
\caption{\small All interlinks of tori derived from six different averagely conservative regions for $\alpha=10$. Here sub-caption $T_{ij}$ denote that the two interlocked tori in the sub-figure are produced by using initial conditions in $D_{i}$ and $D_{j}$ respectively ($1\leq i<j\leq 6$). }
\end{figure}

%Two interlocked tori produced using initial conditions in $D_{5}$ and$D_{6}$

 \indent From Figure 6, it can be seen that the interlinks of tori exhibit many different types. Following [10], we can define \emph{interlinking
 number} to measure the complexity for a interlink of two tori $T_{1}$ and $T_{2}$ in $\mathbb{R}^3$. It can be defined in various ways [11,12], all of which turn out to be equivalent. Here, we adopt the easy way used in [10,13] to clarify the idea which is to count up as follows the crossings of $T_{1}$ and $T_{2}$ in a 'regular projection' of the link (a drawing of it such that no more than two lines cross at any points). Assign orientations and then put an arrow to the two projections of the two tori. Define a number $\epsilon(p)=\pm1$ for 'right' or 'left'-handed crossing $p$ (see Figure 7(a)).Then define the interlinking number as
 \begin{center}
 $Lk(T_{1},T_{2})=\frac{1}{2}|\sum_{p}\epsilon(p)|$.
 \end{center}
 Note that $Lk(T_{1},T_{2})=Lk(T_{2},T_{1})$.
Figure7(b)[10] shows two examples for the definition of interlinking number.
 \begin{figure}[h]
\begin{center}
\subfigure[]{\includegraphics[width=4.2cm]{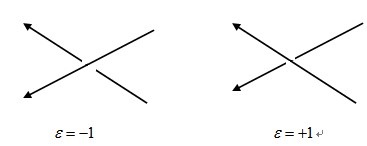}}
\hspace{1cm}
\subfigure[]{\includegraphics[width=4.8cm]{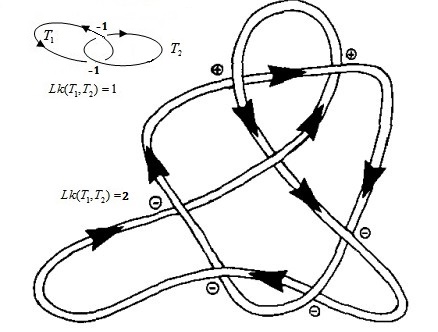}}
\end{center}
\caption{\small (a)Definition of the number $\varepsilon(p)$ for crossing $p$; \small(b) Two examples of interlinking numbers. }
\end{figure}
 % A formal equation is given by Gauss integral []:
 %$Lk(K_{1},K_{2})=-\frac{1}{4}\int_{K_{1}}\int_{K_{2}}\frac{(\emph{\textbf{r}}_1-\emph{\textbf{r}}_2)\cdot d\emph{\textbf{r}}_{1}\times d\emph{\textbf{r}}_{2}}{\|\emph{\textbf{r}}_{1}-\emph{\textbf{r}}_{2}\|}$

\begin{figure}[h]
\begin{center}
\subfigure[$Lk(1,4)=2$]{\includegraphics[width=5.5cm]{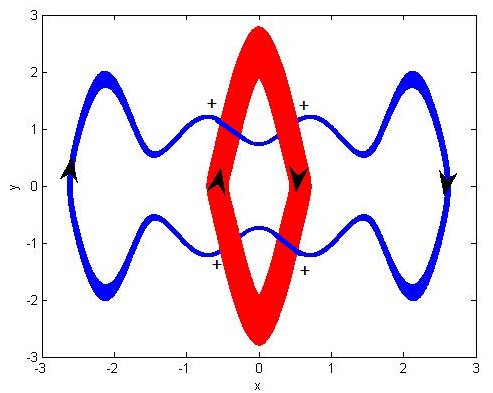}}
\hspace{1cm}
\subfigure[$Lk(4,5)=3$]{\includegraphics[width=5.5cm]{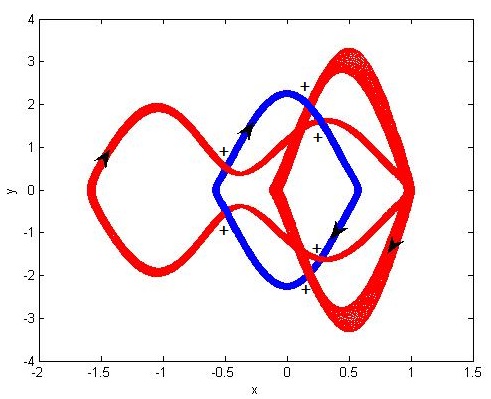}}
\end{center}
\caption{\small The illustrations for the calculations of $Lk(1,4)$ and $Lk(1,2)$.}
\end{figure}
\begin{figure}[h!t]
\centering
\includegraphics[height=3cm]{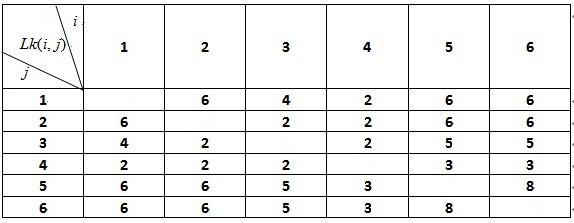}
\renewcommand{\figurename}{Table}
\caption{\small All interlinking numbers $Lk(i,j)$ of the interlinks shown in Figure 6 $(1\leq i\neq j\leq6)$.}
\end{figure}
\indent Now, we calculate all the interlinking numbers of the interlinks of tori shown in Figure 6. We first stress an obvious fact, namely, for any torus $T$ from $D_{i}$ and any torus $\tilde{T}$ from $D_{j}$$(i\neq j)$, $Lk(T,\tilde{T})$ is equivalent. For simplicity, we denote all the same values $Lk(T,\tilde{T})$ by $Lk(i,j)$. According to the above definition of interlinking number, we obtain all thirty values of $Lk(i,j)(1\leq i\neq j\leq6)$ shown in the following Table 9 by choosing applicable projection plane for every different interlinks. Two examples are shown in Figure 8(a) and (b) above for the illustrations of the calculations of $Lk(1,4)$ and $Lk(4,5)$ and both of their projection planes are $xy$-plane.

\section{Conclusions}
This article has investigated the Nos$\acute{\textup{e}}$-Hoover system and reveals that the co-existence of infinite nested tori, a variety of interlinked invariant tori and a chaotic attractor. Furthermore, we have provided a mathematical quantitative description for different interlinks of tori from six different averagely conservative regions by introducing the concept of interlinking number.
It is surprising that the quadratic system which seems so simple possesses so abundantly complicated dynamic properties.\\
\indent It is worth noting that
the system (1), the generalized Nos$\acute{\textup{e}}$-Hoover oscillators studied in [4] and the unusual system studied in [5] are all three-dimensional quadratic systems and have no equilibrium points. Meanwhile, all of them possess invariant tori and chaotic attractors. A natural research topic is when the general quadratic systems without equilibrium points possess the co-existence of invariant tori and chaotic behavior, and moreover possess the interlinked invariant tori. In addition, it is obvious that the interlinking numbers shown in Table 8 will be persisted under small perturbation of parameter $\alpha$ provided that there are no bifurcations involvement. The bifurcation of invariant tori should be another topic to study.

\section{Acknowledgement}
This work was supported by the National Natural Science Foundation of China (11472111) and the Key Disciplines
Construction Foundation of Hefei University (2014xk08).

\textbf{Appendix.} \emph{The Matlab code for the calculation of $det(DP(x_{0}))$(The code for $det(DP^2(x_{0}))$ is similar and we omit it)}\\
function calculate-D \\
options=odeset('RelTol',1e-9,'AbsTol',1e-10,'Events',@events);
format long\\
x0=[-0.72;0;0];
$[$t,x$]$=ode45(@rigid1,[0,1e-6],x0,[]);\\
$[$t,x,te,xe,ie$]$=ode45(@rigid1,[0,300],x(end,:)',options);\\
L=(ie==1);\\
Px0=[xe(L,1),xe(L,3)];\\
x01=[x0(1)+0.000001;0;x0(3)];\\
$[$t,x$]$ = ode45(@rigid1,[0,1e-6],x01,[]);\\
$[$t,x,te,xe,ie$]$=ode45(@rigid1,[0,300],x(end,:)',options);\\
L=(ie==1);\\
Px1=[xe(L,1),xe(L,3)];\\
x02=[x0(1);0;x0(3)+0.000001];\\
$[$t,x$]$ = ode45(@rigid1,[0,1e-6],x02,[]);\\
$[$t,x,te,xe,ie$]$=ode45(@rigid1,[0,300],x(end,:)',options);\\
L=(ie==1);\\
Px2=[xe(L,1),xe(L,3)];\\
J1(:,1)=Px1-Px0;\\
J1(:,2)=Px2-Px0;\\
det(DPx0)=det(1000000*J1);\\

function DX = nose(t,x)\\
DX =[x(2);-x(1)-x(2)*x(3);(x(2)$^2$-1)*10];\\
function [value,isterminal,direction] = events(t,x)\\
value = x(2);\\
isterminal = 1;\\
direction = 1;\\

\end{document}